\documentclass[prb,showpacs,preprintnumbers,amsmath,amssymb,twocolumn,superscriptaddress]{revtex4-1}

\usepackage[dvipdfmx]{graphicx}
\usepackage{dcolumn}
\usepackage{bm}
\usepackage{color}

\usepackage{amsmath}	
\usepackage{amsfonts}

\begin{document}

\title{
Quantum criticality of magnetic catalysis in two-dimensional correlated Dirac fermions
}

\author{Yasuhiro Tada}
\email[]{tada@issp.u-tokyo.ac.jp}
\affiliation{Institute for Solid State Physics, University of Tokyo, Kashiwa 277-8581, Japan}

\begin{abstract}
We study quantum criticality of 
the magnetic field induced charge density wave (CDW) order in correlated spinless Dirac fermions
on the $\pi$-flux square lattice at zero temperature as a prototypical example of the magnetic catalysis,
by using the infinite density matrix renormalization group. 
It is found that the CDW order parameter
$M(B)$ exhibits an anomalous magnetic field $(B)$ scaling behavior characteristic of 
the $(2+1)$-dimensional chiral Ising universality class near the quantum critical point, 
which leads to a strong enhancement of $M(B)$ compared with a mean field result.
We also establish a global phase diagram in the interaction-magnetic field plane 
for the fermionic quantum criticality.
\end{abstract}


\maketitle

\section{introduction}
\label{sec:intro}
Correlated Dirac semimetals are one of the most fundamental systems not only in 
condensed matter physics but also in high energy physics.
They exhibit semimetal-insulator transitions at some critical strength of interactions $V=V_c>0$
at zero temperature,
and magnetic/charge ordered states are stabilized for stronger interactions $V>V_c$
~\cite{Sorella1992,Assaad2013,Wang2014,Wang2016,Li_PRB2015,Li2015,
Hesselmann2016,Huffman2017,Hohenadler2014,Toldin2015,Otsuka2016,Otsuka2018,su4_2018,
Corboz2018,Braun2012,Rosenstein1993,Wetterich2001,Herbut2006,Herbut2009,Herbut2014,Ihrig2018,
DiracQCP}.
These ordered phases correspond to the dynamically massive states with broken chiral symmetry in high energy physics.
Interestingly, criticality of the quantum phase transitions are qualitatively different from those of
conventional magnetic/charge orders in purely bosonic systems,
which is dubbed as fermionic criticality.
In these criticalities, bosonic order parameter fluctuations are intimately coupled with gapless Dirac fermions,
which results in non-trivial quantum critical behaviors depending on fermionic degrees of freedom
in addition to the order parameter symmetry and dimensionality of the system
~\cite{longrange}. 
The fermionic criticality has been discussed extensively by various theoretical methods such as 
lattice model simulations
~\cite{Sorella1992,Assaad2013,Wang2014,Wang2016,Li_PRB2015,Li2015,Hesselmann2016,Huffman2017,
Hohenadler2014,Toldin2015,Otsuka2016,Otsuka2018,su4_2018,
Corboz2018} and
renormalization group calculations
~\cite{Braun2012,Rosenstein1993,Wetterich2001,Herbut2006,Herbut2009,Herbut2014,Ihrig2018},
and now the basic understanding of these systems has been well established.

Correlation effects in a Dirac system become even more significant in presence of an applied 
magnetic field.
It is known that an infinitesimally small magnetic field induces 
a magnetic/charge order for any non-zero interaction $V$, which is called the ``magnetic catalysis"
~\cite{Shovkovy2013book,Miransky2015review,RMP2016,Fukushima2019review,
Gusynin1996,Gusynin1996NPB,Fukushima2012,Scherer2012,QCD1,QCD2,QCD3,QCD4,
graphite2001,Gorbar2002,Semenoff1998,
Roy2008,Roy2011,Roy2014,Boyda2014,DeTar2016,DeTar2017}.
A uniform magnetic field $B$ will effectively reduce spatial dimensionality $d$ of the system via
the Landau quantization, $d\rightarrow d-2$.
Therefore, the system becomes susceptible to formation of a bound state by interactions.
For example in the $(2+1)$-dimensional Gross-Neveu-Yukawa type models, it is shown that in 
the limit of the large number of fermion flavors $N_f$
corresponding to 
a mean field approximation, 
the order parameter behaves as $M(B)\sim B$ for weak interactions $V\ll V_c$,
$M(B)\sim \sqrt{B}$ near the critical point $V=V_c$, and $M(B)-M(0)\sim B^2$ for strong interactions
$V\gg V_c$.
Although the magnetic catalysis was first studied in high energy physics,
it was also discussed in condensed matter physics,
especially for graphene and related materials~\cite{graphite2001,Gorbar2002,Semenoff1998,
Roy2008,Roy2011,Roy2014,Boyda2014,DeTar2016,DeTar2017}.
Recently, there are a variety of candidate Dirac materials with strong electron correlations
~\cite{Hirata2017,CaIrO2019,Sow2017,KondoSemimetal2017,synthetic2011}
which could provide a platform for an experimental realization of the magnetic catalysis,
and a detailed understanding of this phenomenon is an important issue.

However, most of the previous theoretical studies for systems near quantum criticality
are based on perturbative approximations
~\cite{Shovkovy2013book,Miransky2015review,RMP2016,Fukushima2019review,
Gusynin1996,Gusynin1996NPB,Fukushima2012,Scherer2012,QCD1,QCD2,QCD3,QCD4,
graphite2001,Gorbar2002,Semenoff1998,
Roy2008,Roy2011,Roy2014,Boyda2014,DeTar2016,DeTar2017,MCmagcata},
and the true critical behaviors beyond the large $N_f$ limit are 
rather poorly explored.
This is in stark contrast to the correlated Dirac systems without a magnetic field, for which 
there are extensive numerical simulations in addition to the field theoretical calculations,
and critical behaviors have been well established
~\cite{Sorella1992,Assaad2013,Wang2014,Wang2016,Li_PRB2015,Li2015,Hesselmann2016,Huffman2017,
Hohenadler2014,Toldin2015,Otsuka2016,Otsuka2018,su4_2018,Corboz2018,
Braun2012,Rosenstein1993,Wetterich2001,Herbut2006,Herbut2009,Herbut2014,Ihrig2018,DiracQCP}.
Therefore, further theoretical developments are required for clarifying the genuine nature of
the quantum critical magnetic catalysis.

In this work, we study quantum criticality of 
the field induced charge density wave (CDW) order in spinless Dirac fermions
on the two-dimensional $\pi$-flux square lattice, which is one of the simplest realizations of the magnetic catalysis.
We use a non-perturbative numerical technique,
infinite density matrix renormalization group (iDMRG) which can directly describe
spontaneous ${\mathbb Z}_2$ symmetry breaking of the CDW order
~\cite{White1992,DMRG_review1,DMRG_review2,DMRG_review3,TenPy1,TenPy2}.
It is found that the order parameter exhibits an anomalous critical behavior, 
which characterizes the fermionic criticality as clarified by a scaling argument with respect to the magnetic length.
Based on these observations, we establish
a global phase diagram for the ground state near the quantum critical point.

\section{Model}
We consider spinless fermions on a $\pi$-flux square lattice at half-filling under a uniform magnetic field.
There are two Dirac cones in the Brillouin zone and each Dirac fermion has two (sublattice) components,
which corresponds to a case where the total number of Dirac fermion components is four,
similarly to the honeycomb lattice model
~\cite{Wang2014,Wang2016,Li_PRB2015,Li2015,Hesselmann2016,Huffman2017}.
The Hamiltonian is given by
\begin{align}
H=-\sum_{\langle i,j\rangle}t_{ij}c^{\dagger}_ic_j+V\sum_{\langle i,j\rangle}n_in_j,
\label{eq:H}
\end{align}
where $\langle i,j\rangle$ is a pair of the nearest neibghbor sites and the energy unit is $t=1$.
The hopping is $t_{ij}=te^{i\pi y_i}\exp (iA_{ij})$ 
along the $x$-direction on the $y=y_i$ bond and $t_{ij}=t\exp (iA_{ij})$ along the $y$-direction. 
The vector potential is given in the string gauge~\cite{Hatsugai1999} with the period $L_x'\times L_y$ 
where $L_x'$ is 
the superlattice unit period used in the iDMRG calculations for the system size $L_x\times L_y=\infty\times L_y$.
Typically, we use $L_x'=20$ for $L_y=6$ and $L_x'=10$ for $L_y=10$.
See also Appendix~\ref{app:iDMRG}.
$A_{ij}=0$ corresponds to the conventional $\pi$-flux square lattice without an applied field,
while $A_{ij}\neq0$ describes
an applied magnetic field for a plaquette $p$, $B_p=\sum_{\langle ij\rangle\in p}A_{ij}$.
The magnetic field is spatially uniform and an integer multiple 
of a unit value $B=n\times \delta B \quad (n=1,2,\cdots, L_x'L_y)$ allowed by the superlattice size, where 
$\delta B=2\pi/L_x'L_y$.
The lattice constant $a$ as a length unit and the electric charge $e$ have been set as $a=1,e=1$,
and the magnetic field is measured in the unit of $B_0=2\pi$.

The $V$-term 
is a repulsive nearest neighbor interaction leading to the CDW state and the quantum phase transition with $B=0$
takes place at $V=V_c\simeq 1.30t$ according to the quantum Monte Carlo calculations for 
the bulk two dimensional system, where 
the criticality belongs to the $(2+1)$-dimensional chiral Ising universality class~\cite{Wang2014,Wang2016,Li_PRB2015,Li2015,Hesselmann2016,Huffman2017}.
On the other hand, our cylinder system is anisotropic and the CDW quantum phase transition
at $B=0$ is simply (1+1)-dimensional Ising transition and critical interaction strength depends 
on the system size $L_y$, which may be regarded as a finite size effect~\cite{Tada2019}.
However, the system can be essentially two-dimensional in space under a magnetic field
when the magnetic length $l_B=1/\sqrt{B}$ becomes shorter than the system size $L_y$.
We will use this property to discuss the $(2+1)$-dimensional criticality. 
Note that the critical interaction strength $V_c\simeq 1.30t$ will be confirmed later within the present framework.

In the following, we focus on the CDW order parameter,
\begin{align}
M=\left| \frac{1}{L_x'L_y}\sum_i(-1)^{|i|}n_i \right|,
\end{align} 
where the summation runs over the superlattice unit cell.
In the iDMRG calculation, we introduce a finite bond dimension $\chi$ up to $\chi=1600$ as a cut-off
to approximate the ground state wavefunction in the form of a matrix product state,
and we can obtain the true ground state by a careful extrapolation to
$\chi\rightarrow\infty$ from the finite $\chi$ results
~\cite{White1992,DMRG_review1,DMRG_review2,DMRG_review3,TenPy1,TenPy2}
(see also Appendix~\ref{app:iDMRG}).
Generally, iDMRG with finite $\chi$ gives a good approximation especially when the system considered 
is gapful.
As we will show, an extrapolation to $\chi\rightarrow\infty$ works well, 
because our system has a gap in presence of
a non-zero $B$ due to the magnetic catalysis of the broken discrete symmetry ${\mathbb Z}_2$ 
where there is no gapless Nambu-Goldstone mode. 
For a comparison,
we also discuss a two-leg ladder system with $L_y=2$ in Appendix~\ref{app:a}.

\section{Away from quantum critical point}
\label{sec:nonQCP}
\begin{figure}[tbh]
\includegraphics[width=5.5cm]{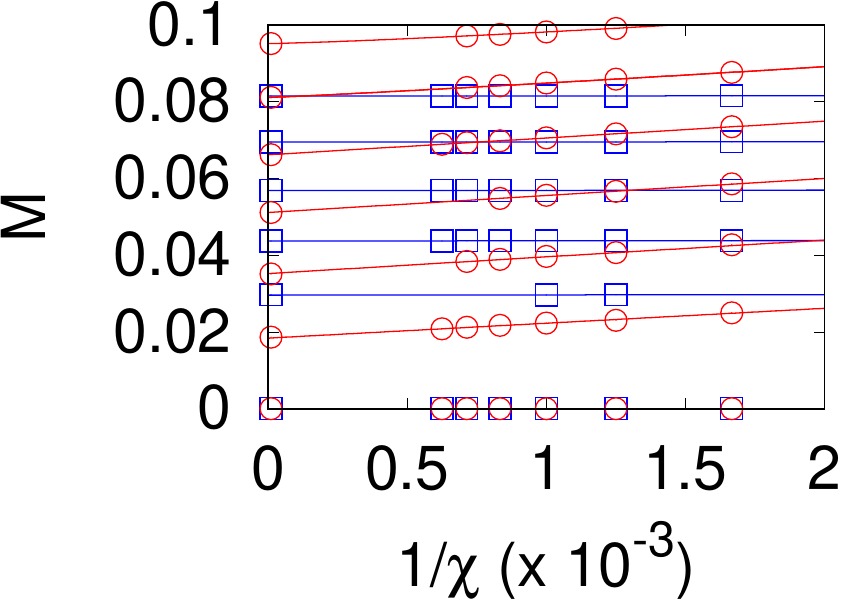}
\caption{Extrapolation of the CDW order parameter $M(B)$ for the $\chi\rightarrow\infty$ limit
at $V=0.50t$. 
The blue (red) symbols are for $L_y=6 (L_y=10)$ and the curves are power law fittings.
Each curve corresponds to a magnetic field in the range $0\leq B\leq 0.06B_0$.
}
\label{fig:extrap}
\end{figure}

Before discussing quantum criticality, 
we investigate the magnetic catalysis when the system is away from the critical point. 
Firstly, we consider a weak interaction $V=0.50t<V_c=1.30t$ for which the system at $B=0$ is a Dirac semimetal
renormalized by the interaction. 
As exemplified in Fig.~\ref{fig:extrap}, dependence of $M(B)$ on the bond dimension $\chi$ used in the calculation 
is negligibly small for $L_y=6$, and it can be safely extrapolated to $\chi\rightarrow\infty$ even for 
$L_y=10$.
Standard deviations of the extrapolations are less than 1\% and within the symbols.
Such an extrapolation can be done also for other values of $V$ as mentioned before,
and all results shown in this study are extrapolated ones. 

\begin{figure}[tbh]
\includegraphics[width=5.5cm]{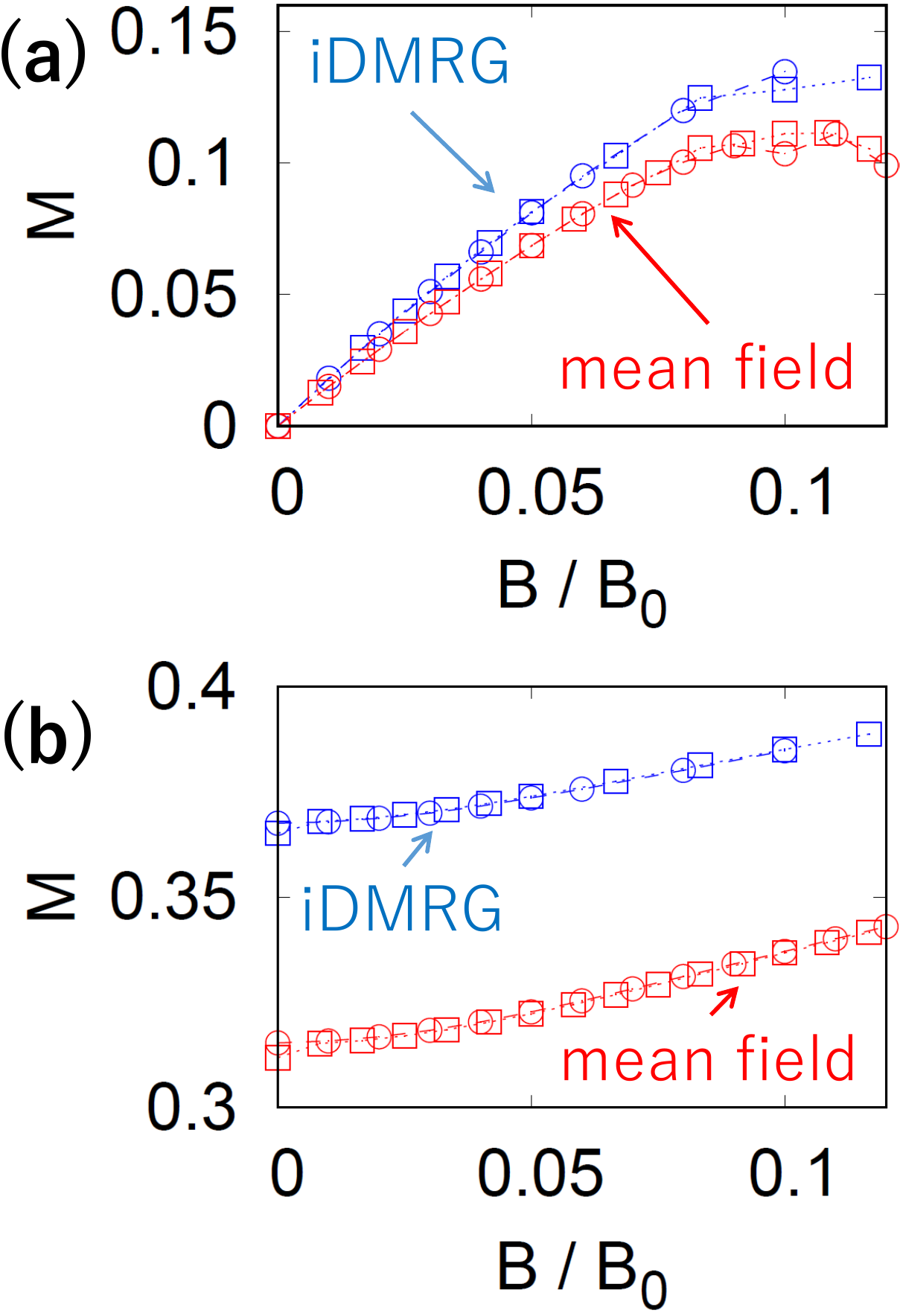}
\caption{
(a) The CDW order parameter $M$ at a weak coupling. 
The blue symbols are the iDMRG results at $V=0.50t<V_c$ for 
$L_y=6$ (squares) and $L_y=10$ (circles), 
while the red symbols are the mean field results ($V=0.30t$) for the same system sizes.
(b) $M$ at a strong coupling $V=2.0t>V_c$ calculated by iDMRG (blue) and 
$V=1.20t$ by the mean field approximation (red).
The interactions for iDMRG and the mean field approximation correspond to
the same value of the normalized coupling constant $g$.  
}
\label{fig:MV052}
\end{figure}

In Fig.~\ref{fig:MV052} (a), we show the CDW order parameter $M$ extrapolated to $\chi\rightarrow\infty$
for the system sizes $L_y=6$ and $L_y=10$
at $V=0.50t$. 
The calculated results almost converge for $L_y=6,10$ and are independent of the system size,
except for $B=0$ where there is a finite size effect due to $l_B=\infty$,
although there is some accidental deviation around $B\simeq0.1B_0$. 
Therefore, these results give the CDW order parameter essentially in the thermodynamic limit $L_y\rightarrow\infty$.
In order to understand impacts of quantum fluctuations,
we also performed a mean field calculation for a comparison~\cite{MF}.
The critical interaction within the mean field approximation is found to be $V_c=0.78t$
and the interaction $V=0.30t$ corresponds to
the same coupling strength in terms of the normalized interaction $g=(V-V_c)/V_c=-0.62$. 
The iDMRG reuslts of $M$ (blue symbols) are larger than the corresponding mean field results (red symbols), 
$M>M_\textrm{MF}$,
which suggests that quantum fluctuations enhance the order parameter even for the present weak $V$.
It is noted that the order parameter behaves roughly as $M(B)\sim B$ as seen for small $B$, 
which is consistent with the large $N_f$ field theory
~\cite{Shovkovy2013book,Miransky2015review,RMP2016,Fukushima2019review,
Gusynin1996,Gusynin1996NPB,Fukushima2012,Scherer2012}.

Similarly to the weak interaction case, the CDW order parameter $M$ calculated by iDMRG (blue symbols) 
is enhanced at 
a strong interaction $V=2.0t>V_c$ compared to the mean field result $M_\textrm{MF}$ (red symbols) 
at the corresponding interaction
$V=1.20t$ (or equivalently $g= 0.54$) as shown in Fig.~\ref{fig:MV052} (b).
However, this is governed by the $B=0$ values and increase of $M(B)$ by the magnetic field
is roughly comparable to that of $M_\textrm{MF}(B)$.
The result that $M>M_\textrm{MF}$ already at $B=0$ is because they behave as $M(V,B=0)\sim g^{\beta}$ with
$\beta\simeq 0.5\sim 0.6<1$~\cite{Wang2014,Wang2016,Li_PRB2015,Li2015,Hesselmann2016,Huffman2017}  
while $M_\textrm{MF}(V,B=0)\sim g^{\beta_\textrm{MF}}$ with $\beta_\textrm{MF}=1$ near the quantum critical point,
and these critical behaviors essentially determine magnitudes of 
the CDW order parameters away from the critical points.
For $B\neq 0$, the order parameter behaves roughly as $M(B)-M(0)\sim B^2$ in agreement with the
large $N_f$ field theory~\cite{Shovkovy2013book,Miransky2015review,RMP2016,Fukushima2019review,
Gusynin1996,Gusynin1996NPB,Fukushima2012,Scherer2012}.

\section{Near quantum critical point}
In this section, we discuss quantum criticality of the magnetic catalysis based on a variant of 
finite size scaling ansatzes.
Then, we establish a global phase diagram around the quantum critical point 
in the interaction-magnetic field plane,
in close analogy with
the well-known finite temperature phase diagram near a quantum critical point.

\subsection{Scaling argument}

The enhancement of $M(B)$ by the quantum fluctuations can be 
even more pronounced
near the quantum critical point. 
\begin{figure}[tbh]
\includegraphics[width=5.5cm]{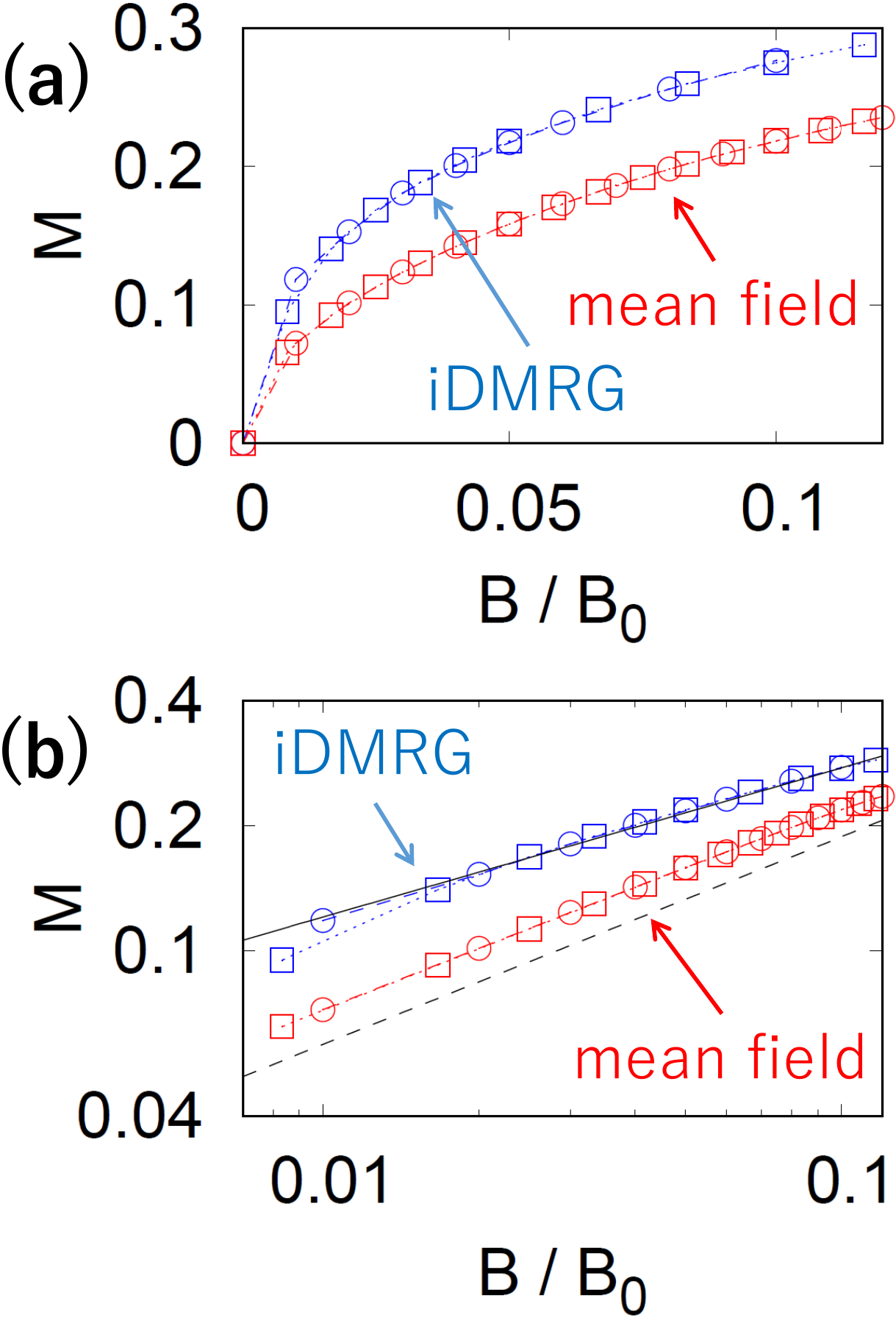}
\caption{
(a) The CDW order parameter $M$ at the quantum critical point $V=V_c=1.30t$ calculated by iDMRG
together with the mean field result at $V=0.78t$ corresponding to $g=0$. 
Definitions of the symbols are the same as in Fig.~\ref{fig:MV052}.
(b) $M$ in the log-log plot. The black solid line is the power law fitting $M\sim B^{0.355}$, while
the black dashed line is the large $N_f$ result $M\sim\sqrt{B}$ shown for the eyes.
}
\label{fig:MV13}
\end{figure}
Figure~\ref{fig:MV13} shows the CDW order parameter at $V=V_c=1.30t$ (blue symbols) 
together with the mean field result
for $V=0.78t$ (red symbols), corresponding to $g= 0$.
Clearly, the iDMRG result is significantly larger than the mean field result,
and the enhancement is much stronger than that in the weak interaction case. 
There are some deviations between the results for $L_y=6$ and $L_y=10$ for small magnetic fields,
$B\lesssim 0.01B_0$, due to a long magnetic length $l_B$,
and the CDW order gets more strongly stabilized when the system size $L_y$ increases from $L_y=6$ to $L_y=10$.
This should be a general tendency 
since the CDW phase at $B=0$ extends to a smaller interaction region when the system size
increases~\cite{Tada2019}.
From this observation, we can discuss scaling behaviors of the CDW order parameter 
in the thermodynamic limit as
a function of $B$ near the quantum critical point.
Indeed, as shown in Fig.~\ref{fig:MV13} (b), the calculated $M$ except for the smallest values of $B$ converge
for different system sizes $L_y=6, 10$, and $M(B)$ exhibits a power law behavior
for $0.02B_0\lesssim B\lesssim 0.1B_0$. 
The finite size effects are negligible in this range of the magnetic field,
and furthermore the scaling behavior would hold for smaller magnetic fields down to $B=0$ 
in a thermodynamic system $L_y\rightarrow\infty$, since $M(L_y=10)$ shows the scaling behavior
in a wider region of $B$ than $M(L_y=6)$ does.
If we focus on $0.02B_0\lesssim B\lesssim 0.1B_0$ in Fig.~\ref{fig:MV13},
we obtain the anomalous scaling behavior
$M(B)\sim B^{0.355(6)}$ by power law fittings for different sets of data points.
This is qualitatively different from the mean field (or equivalently large $N_f$ limit) result $M_\textrm{MF}\sim \sqrt{B}$,
which eventually leads to the strong enhancement of $M(B)$ compared to $M_\textrm{MF}(B)$.

The calculated magnetic field dependence of the CDW order parameter
near $V= V_c$ implies a scaling relation characteristic of the quantum criticality.
Here, we propose a scaling ansatz for the leading singular part of 
the ground state energy density of a thermodynamically large $(2+1)$-dimensional system,
\begin{align}
\varepsilon_\textrm{sing}(g,h,l_B^{-1})=b^{-D}\varepsilon_\textrm{sing}(b^{y_g}g,b^{y_h}h,bl_B^{-1}),
\label{eq:ansatz}
\end{align} 
where $D=2+z=2+1=3$ with $z=1$ being the dynamical critical exponent and $h$ is the conjugate field 
to the CDW order parameter $M$. 
The exponents $y_{g,h}$ are corresponding scaling dimensions, and 
the scaling dimension of the magnetic length is assumed to be one as will be confirmed later.
For a thermodynamic system, the magnetic length $l_B$ will play a role of
a characteristic length scale similarly to a finite system size $L$.
Then, a standard argument similar to that for a finite size system at $B=0$ applies, leading to 
\begin{align}
M(g=0,l_B^{-1})\sim (l_B^{-1})^{\beta/\nu}\sim B^{\beta/2\nu},
\label{eq:M0}
\end{align} 
where $\beta$ and $\nu$ are the critical exponents at $B=0$ for the order parameter
$M(g,l_B^{-1}=0)\sim g^{\beta}$ and the correlation length $\xi(g,l_B^{-1}=0)\sim g^{-\nu}$.
One sees that this coincides with the familiar finite size scaling if we replace $l_B$ with 
a system size $L$~\cite{Cardy}.
The critical exponents of the CDW quantum phase transition in $(2+1)$-dimensions 
are $\beta=\nu=1$ in the mean field approximation,
and the resulting $M\sim B^{0.5}$ is consistent with our mean field numerical calculations
~\cite{hyperscaling}.
The true 
critical exponents for the present $(2+1)$-dimensional chiral Ising universality class 
with four Dirac fermion components
have been obtained by the quantum Monte Carlo simulations at $B=0$,
and are given by $(\beta=0.53, \nu=0.80)$~\cite{Wang2014,Wang2016}, 
which was further supported by the infinite projected entangled pair state calculation~\cite{Corboz2018}.
Other quantum Monte Carlo studies with different schemes and system sizes
give $(\beta=0.63, \nu=0.78)$~\cite{Li_PRB2015,Li2015},
$(\beta=0.47,\nu=0.74)$~\cite{Hesselmann2016}, and $(\beta=0.67,\nu=0.88)$~\cite{Huffman2017}.
These exponents lead to $\beta/2\nu=0.33, 0.40, 0.32, 0.38$ respectively, 
and the scaling behavior of $M(B)$ found in our study 
falls into this range and is consistent with them.

\begin{figure}[tbh]
\includegraphics[width=6.5cm]{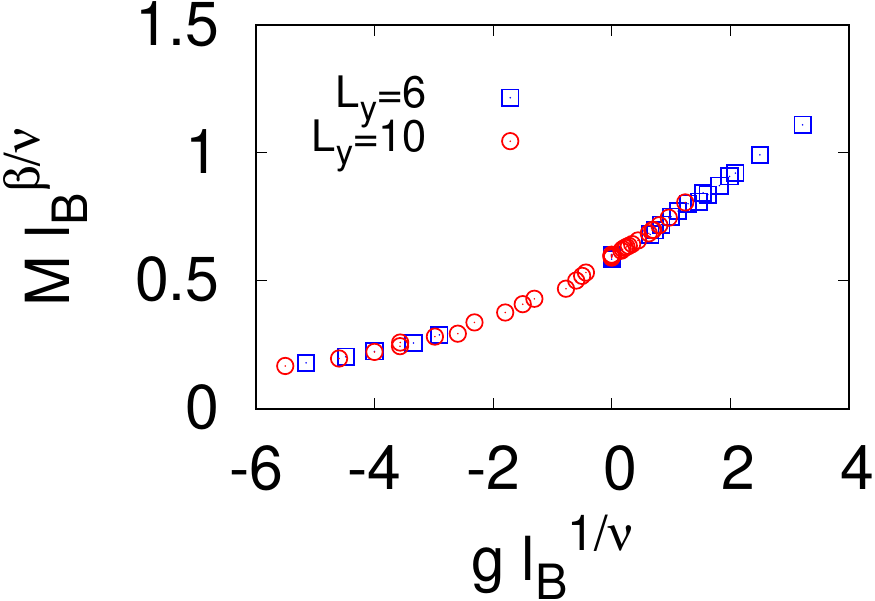}
\caption{Scaling plot of the CDW order parameter $M(V,B)$ in terms of $g=(V-V_c)/V_c$ and $l_B=1/\sqrt{B}$.
The blue squares are for $L_y=6$ and red circles for $L_y=10$.
}
\label{fig:scaling}
\end{figure}

The homogeneity relation Eq.~\eqref{eq:ansatz} and the critical exponent can be further confirmed 
by performing a data collapse.
According to Eq.~\eqref{eq:ansatz},
the CDW order parameter for general $g$ is expected to behave as
\begin{align}
M(g,l_B^{-1})= l_B^{-\beta/\nu}\Phi(gl_B^{1/\nu}),
\end{align} 
where $\Phi(\cdot)$ is a scaling function.
This is a variant of the finite size scaling similarly to Eq.~\eqref{eq:M0}.
When performing a data collapse,
we use the results for $0.02B_0\lesssim B\lesssim 0.1B_0$
so that finite size effects are negligible.
As shown in Fig.~\ref{fig:scaling}, 
the calculated data well
collapse into a single curve and the critical exponents are evaluated as 
$\beta=0.54(3),\nu=0.80(2)$ with $V_c=1.30(2)t$.
This gives $\beta/2\nu= 0.34(2)$, which
is consistent with $\beta/2\nu=0.36$ obained from $M(V=V_c,B)$ at the quantum critical point (Fig.~\ref{fig:MV13}).
Our critical exponents are compatible with those obtained previously by the numerical calculations as mentioned above
and roughly with those by the field theoretic calculations
~\cite{Sorella1992,Assaad2013,Wang2014,Wang2016,Li_PRB2015,Li2015,
Hesselmann2016,Huffman2017,Hohenadler2014,Toldin2015,Otsuka2016,Otsuka2018,su4_2018,
Corboz2018,Braun2012,Rosenstein1993,Wetterich2001,Herbut2006,Herbut2009,Herbut2014,Ihrig2018,DiracQCP}.
Our numerical calculations for the $(2+1)$-dimensional criticality
are limited to rather small magnetc lengths $l_B$ bounded by the system size $L_y$,
and we expect that more accurate evaluations of the critical exponents would be possible
for larger $L_y$ with controlled extrapolations to $\chi\rightarrow \infty$.

The successful evaluation of the critical exponents 
strongly verifies the scaling ansatz Eq.~\eqref{eq:ansatz}. 
Although the scaling ansatz may be intuitively clear and similar relations were discussed for the bosonic 
Ginzburg-Landau-Wilson theory in the context of the cuprate high-$T_c$ superconductivity
~\cite{Fisher1991,Lawrie1997,Tesanovic1999},
its validity is {\it a priori} non-trivial and 
there have been no non-perturbative analyses even for the well-known bosonic criticality.
This is in stark contrast to the conventional finite system size scaling at $B=0$ which
has been well established for various systems~\cite{Cardy}.
The present study is a first non-perturbative analysis of the $l_B$-scaling relation,
providing a clear insight from a statistical physics point of view for the quantum critical magnetic catalysis.
Besides, the scaling ansatz could be used as a theoretical tool for investigating some critical phenomena similarly to
the recently developed finite correlation length scaling in tensor network states
(see also Appendix~\ref{app:a})~\cite{Corboz2018,Tada2019,Rader2018,Pollmann2009}.
Based on this observation,
one could evaluate critical behaviors of the magnetic catalysis in other universality classes in $(2+1)$-dimensions,
such as $\mathrm{SU}(2)$ and $\mathrm{U}(1)$ symmetry breaking with a general number of Dirac fermion components,
by using the critical exponents obtained for the phase transitions at $B=0$
~\cite{Sorella1992,Assaad2013,Wang2014,Wang2016,Li_PRB2015,Li2015,
Hesselmann2016,Huffman2017,Hohenadler2014,Toldin2015,Otsuka2016,Otsuka2018,su4_2018,
Corboz2018,Braun2012,Rosenstein1993,Wetterich2001,Herbut2006,Herbut2009,Herbut2014,Ihrig2018,DiracQCP}.
It would be a future problem to clarify
the exact condition for the $l_B$-scaling to hold in general cases.

\subsection{Phase diagram}

The above discussions can be summarized into a global phase diagram near the quantum critical point
in the $V$-$B$ plane at zero temperature 
as shown in Fig.~\ref{fig:scaling}.
Here, we mainly focus on the critical behaviors of the order parameter but not on phase boundaries.
In this phase diagram, there are two length scales; one is the correlation length of the CDW order parameter $\xi$
at $B=0$,
and the other is the magnetic length $l_B$. 
One can compare it with the familiar finite temperature phase diagram near a quantum critical point
~\cite{QCPreview1997,MoriyaUeda2000,QCPreview2007}.
The length scale $l_B$ in our case corresponds to
a system size along the imaginary time, $L_{\tau}=1/T$, 
in a standard quantum critical system at finite temperature $T$.
In a finite temperature system,  anomalous finite temperature behaviors are seen when
the dynamical correlation length $\xi_{\tau}\sim \xi^z$ becomes longer than the temporal 
system size, $\xi_{\tau}\gg L_{\tau}$, so that the critical singularity is cut off by $L_{\tau}$ in the imaginary time
direction~\cite{QCPreview1997,MoriyaUeda2000,QCPreview2007}.
Similarly in the present system at $T=0$,
physical quantities will exhibit anomalous $B$-dependence 
when the spatial correlation length $\xi$ is longer than the magnetic length, $\xi\gg l_B$,
and the critical singularity is cut off by $l_B$ in the spatial direction.  
In this way, we can understand the scaling behavior $M\sim l_B^{-\beta/\nu}\sim B^{\beta/2\nu}$ in close analogy with
the finite temperature scaling behaviors associated with a quantum critical point at $B=0$.
On the other hand, the order parameter shows conventional $B$-dependence, 
$M(B)\sim B$ or $M(B)-M(0)\sim B^2$, when the system is away from the quantum critical point, 
$\xi\ll l_B$.
We note that our phase diagram would be qualitatively applicable to an interacting Dirac system 
with a general flavor number $N_f$
including $N_f\rightarrow\infty$ with $\beta=\nu=1$~\cite{Shovkovy2013book,Miransky2015review}.
It is also noted that the Dirac semimetal phase will be extended to a $B\neq 0$ region at finite low temperature
~\cite{Shovkovy2013book,Miransky2015review,QCD1,QCD2,QCD3,QCD4,Boyda2014,DeTar2016,DeTar2017},
and the critical behaviors can be modified as will be briefly discussed later.
\begin{figure}[tbh]
\includegraphics[width=5.5cm]{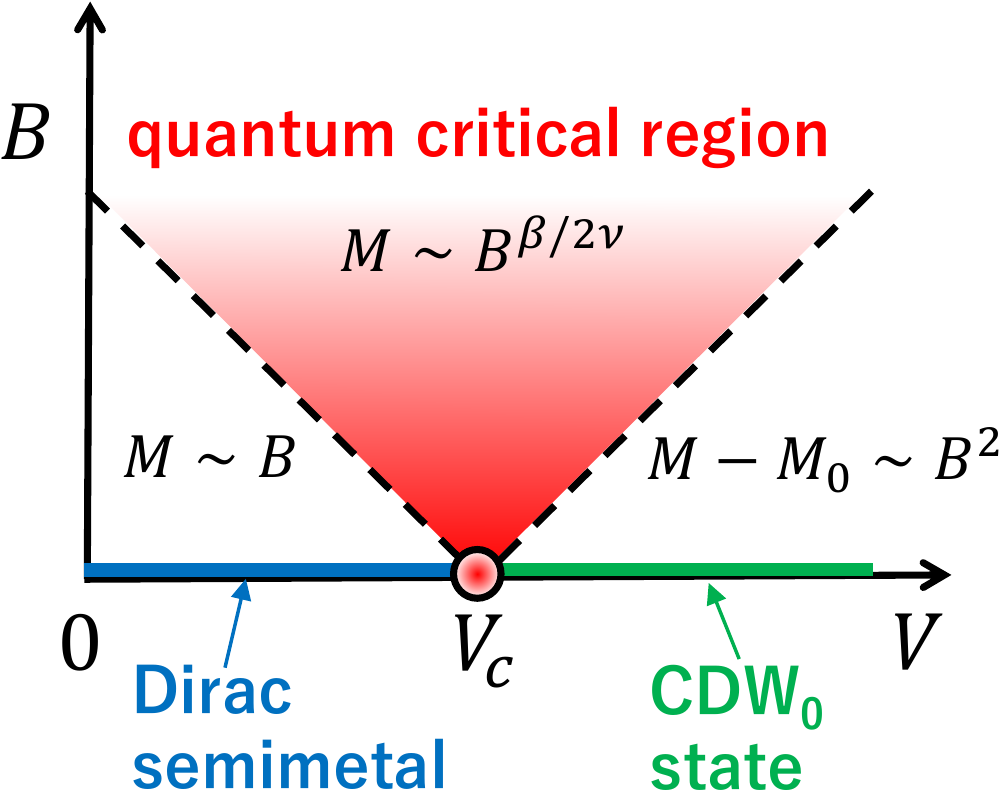}
\caption{Schematic phase diagram in the $V$-$B$ plane at zero temperature
and the $B$-dependence of $M(V,B)$ for fixed $V$ in each region.
The CDW state at $B=0$ is denoted as CDW$_0$ and 
$M_0(V)=M(V,B=0)\sim(V-V_c)^{\beta}$.
The crossover boundaries (dashed lines) are characterized by
$l_B\simeq \xi$.
}
\label{fig:phase_diagram}
\end{figure}

We would also expect that a similar phase diagram could be seen even in a system with long-range interactions such as
QED-like theories in the massless limit,
because it is considered that criticality of a quantum phase transition in a $(2+1)$-dimensional Dirac system 
driven by a short-range interaction
is not affected by the long-range Coulomb interaction~\cite{Hohenadler2014,Herbut2009}.
It is noted that, while the Coulomb interaction is (marginally) irrelevant at the transition point,
it will play an important role at a weak coupling regime and an order parameter
could behave as $M\sim \sqrt{B}$ even for any small coupling~\cite{Shovkovy2013book,Miransky2015review}.

\subsection{Discussions}

In this section, we discuss several issues in the magnetic catalysis including possible future studies.

{\it Comparison with conventional finite size effects}---
In the previous section, we have discussed the effects of a finite $l_B$ in analogy with 
the temporal size $L_{\tau}$.
Here, we make a comparison of the magnetic catalysis as a finite size effect in spatial directions
with the conventional finite size effects.
In a finite size Dirac system with an isotropic linear system size $L$ in absence of a magnetic field, 
an order parameter $M$ (more precisely, a long range order $M=\sqrt{\langle \hat{M}^2\rangle}$) 
is usually overestimated when compared with the thermodynamic value,
and it shows smooth crossover for a wide range of interaction strength when the system size is fixed
~\cite{Sorella1992,Assaad2013,Wang2014,Wang2016,Li_PRB2015,Li2015,Hesselmann2016,Huffman2017,
Hohenadler2014,Toldin2015,Otsuka2016,Otsuka2018,su4_2018}.
For different system sizes, it behaves as $M\sim L^{-\beta/\nu}$ at the critical point based on the
conventional finite size scaling ansatz.
Similar scaling relations hold also for an infinite system within a framework of tensor network states where
the system size $L$ is replaced by the correlation length due to a finite bond dimension
(see also Appendix~\ref{app:a})~\cite{Corboz2018,Tada2019}.
In this sense, at least formally, 
the enhanced $M$ by the magnetic field in the present study is analogous
to the overestimated $M$ in a conventional finite size system without a magnetic field.
Furthermore, these two phenomena share a physical origin in common, i.e. the dimensional reduction.
As mentioned in Sec.~\ref{sec:intro},
a magnetic field reduces the spatial dimensionality $d\rightarrow d-2$ via the Landau quantization.
Similarly, a small system size quantizes the spatial degrees of freedom and possible
wavenumbers are discretized.
Consequently, the density of states at low energy can become larger than that in the thermodynamic limit
and correlation effects can be amplified,
which would lead to enhanced/overestimated $M$.
Therefore, the magnetic catalysis can be regarded as a finite size effect and is expected to be a quite universal
phenomenon.
However,
there is a crucial difference that the finite $l_B$ effect can be observed in an experiment as 
an anomalous $B$-dependence $M(B)\sim l_B^{-\beta/\nu}
\sim B^{\beta/2\nu}$,
in contrast to the familiar finite size scaling, $M\sim L^{-\beta/\nu}$.

{\it Ground state energy density}---
Although we have been focusing on the CDW order parameter,
scaling behaviors will also be seen in other quantities such as the ground state energy density $\varepsilon$ itself.
According to Eq.~\eqref{eq:ansatz}, $\varepsilon$ of a thermodynamically large system is expected to behave as 
\begin{align}
\varepsilon(g,l_B^{-1})=\varepsilon(g,0) +\frac{\varepsilon_\textrm{sing}(g l_B^{1/\nu})}{l_B^3}+\cdots.
\end{align} 
At the quantum critical point $g=0$ (i.e. $V=V_c$), 
the prefactor in front of $l_B^{-3}$ might be factorized as $\varepsilon_\textrm{sing}(0)=C_0v$ with a constant $C_0$
and the ``speed of light" $v$ characterizing the underlying field theory with
the Lorentz invariance
~\cite{Rader2018}.
Away from the quantum critical point, the mean field behaviors will be qualitatively correct 
as we have seen in the CDW order parameter $M$ (Sec.~\ref{sec:nonQCP}).
Indeed, our iDMRG calculation and mean field calculation suggest
for a small magnetic field $l_B^{-1}\rightarrow0$,
$\varepsilon_\textrm{sing}(g l_B^{1/\nu}\ll -1)\sim $ const $>0$
in the Dirac semimetal regime $g<0$ (i.e. $V< V_c$), 
while $\varepsilon_\textrm{sing}(g l_B^{1/\nu}\gg 1)\sim l_B^{-1}>0$ in the ordered phase $g>0$ (i.e. $V> V_c$),
which is in agreement with the large $N_f$ field theory~\cite{Shovkovy2013book,Miransky2015review}.
Consequently, the orbital magnetic moment $m_\textrm{orb}=-\partial \varepsilon/\partial B$ will be 
$m_\textrm{orb}\sim -\sqrt{B}$ for the former (and also at the critical point), and 
$m_\textrm{orb}\sim -B$ for the latter.
Details of the ground state energy density and the diamagnetic orbital magentic moment will be discussed elsewhere.

{\it Finite temperature correction}---
Finally, we briefly touch on finite temperature effects around $T= 0$.
At finite temperature, the new length scale $L_{\tau}$ is introduced and
we expect an anomalous $T/\sqrt{B}$ scaling in our system,
by following a scaling hypothesis for the singular part of the free energy density,
$f_\textrm{sing}(g,h,l_B^{-1},L_{\tau}^{-1})=b^{-D}f_\textrm{sing}(b^{y_g}g,b^{y_h}h,bl_B^{-1},b^zL_{\tau}^{-1})$ with $z=1$.
For example, the CDW order parameter would have a finite temperature correction given by
$M(B,T)= B^{\beta/2\nu}\Psi(T/\sqrt{B})$ at the critical point $g=0$, 
where $\Psi(\cdot)$ is a scaling function with the property $\Psi(x\rightarrow0)=$ const.
Since finite temperature effects are important in experiments,
detailed investigations of them would be an interesting future problem. 

\section{summary}
We have discussed quantum criticality of the magnetic catalysis
in spinless fermions
on the $\pi$-flux square lattice by non-perturbative calculations with iDMRG.
We found the scaling behavior of the CDW order parameter $M(B)$ 
characteristic of the $(2+1)$-dimensional chiral Ising
universality class, and established a global phase 
diagram near the quantum critical point.
The present study is a first non-perturbative investigation of fermionic quantum criticality under a magnetic field, 
and could provide a firm basis for deeper understandings of other related systems.

\section*{acknowledgements}
We thank F. Pollmann for introducing the open source code TenPy for the iDMRG calculations.
We are also grateful to Y. Fuji, M. Oshikawa, and K. Fukushima for valuable discussions.
The numerical calculations were performed at Max Planck Institute for the Physics of Complex Systems. 
This work was supported by JSPS KAKENHI Grant No. JP17J05736,
No. JP17K14333, KAKENHI on Innovative Areas ``J-Physics''
[No. JP18H04318].

\appendix
\section{Quick overview of iDMRG}
\label{app:iDMRG}
In this section, we briefly touch on the basics of iDMRG~\cite{White1992,DMRG_review1,DMRG_review2,DMRG_review3,TenPy1,TenPy2}.
The DMRG is a variational method to accurately simulate a target quantum state based on
the framework of matrix product states.
A ground state in a one-dimensional system can be expressed in the form of a matrix product state,
\begin{align}
|\Psi\rangle = \sum_{i_1,\cdots,i_N} \textrm{Tr}[M^{[1]i_1}\cdots M^{[N]i_N}]|i_1,\cdots,i_N\rangle,
\end{align} 
where $N$ is the system size and $\{|i_1,\cdots,i_N\rangle\}$ is a local basis.
The matrix $M$ is decomposed to $M=U\Lambda V^{\dagger}$
by the singular value decomposition,
and only largest $\chi$ 
singular values in the diagonal matirx $\Lambda$ are kept in numerical calculations.
This works quite well particularly for a gapped system where the singular values decay exponentially in $\chi$.
The optimal matrices are found by minimizing the variational state energy.

In iDMRG, we assume that the matrices $\{M^{[k]}\}$ have a periodicity $N'$,
such that $M^{[k]}=M^{[k+N']}$.
This enables us to formally consider an infinitely large system by repeating the unit cell structure,
\begin{align}
|\Psi\rangle &= \sum_{} \textrm{Tr}[\cdots M^{[N']i_{0}}M^{[1]i_1}\cdots M^{[N']i_{N'}}M^{[1]i_{N'+1}}\cdots]\nonumber\\
&\quad \times|\cdots,i_0,i_1,\cdots,i_{N'},i_{N'+1},\cdots\rangle.
\end{align} 

One can also use this scheme to study a two-dimensional system by 
introducing a ``snake-like trace" of the two-dimensional lattice and regarding it
as a one-dimensional system with long-range hopping/interactions.
In our study, we consider $L_x\times L_y=\infty\times L_y$ system with the period $L_x'$ along the $x$-direction.
This system can be regarded as an infinite one-dimensional system with the period 
$N'=L_x'\times L_y$, and such a one-dimensional system can be described by a matrix product state.
Detailed discussions and applications can be found in the literature~\cite{White1992,DMRG_review1,DMRG_review2,DMRG_review3,TenPy1,TenPy2}.

\section{Two-leg ladder}
\label{app:a}
We briefly discuss numerical results for a two-leg ladder system at half-filling for a comparison.
Here, we do not use the $\chi\rightarrow\infty$ extrapolation but instead apply
correlation length scaling for several values of $\chi$.
The two-leg ladder system has been extensively investigated with and without magnetic fields
~\cite{Tada2019,Carr2006},
but criticality of a magnetic field induced order has not been examined before.
We consider the Hamiltonian Eq.~\eqref{eq:H} where, instead of the string gauge, the hopping integrals are now
$t_{ij}=te^{i\phi/2}$ along the chain-1, $t_{ij}=-te^{-i\phi/2}$ along the chain-2, and 
the inter-chain hopping $t_{ij}=t$.
This realizes a magnetic field $B=\phi/a^2$ with the lattice constant $a=1$,
but note that the magnetic length plays no role in the present two-leg ladder since the
system size in the $y$-direction is only $L_y=2$.

\begin{figure}[tbh]
\includegraphics[width=5.5cm]{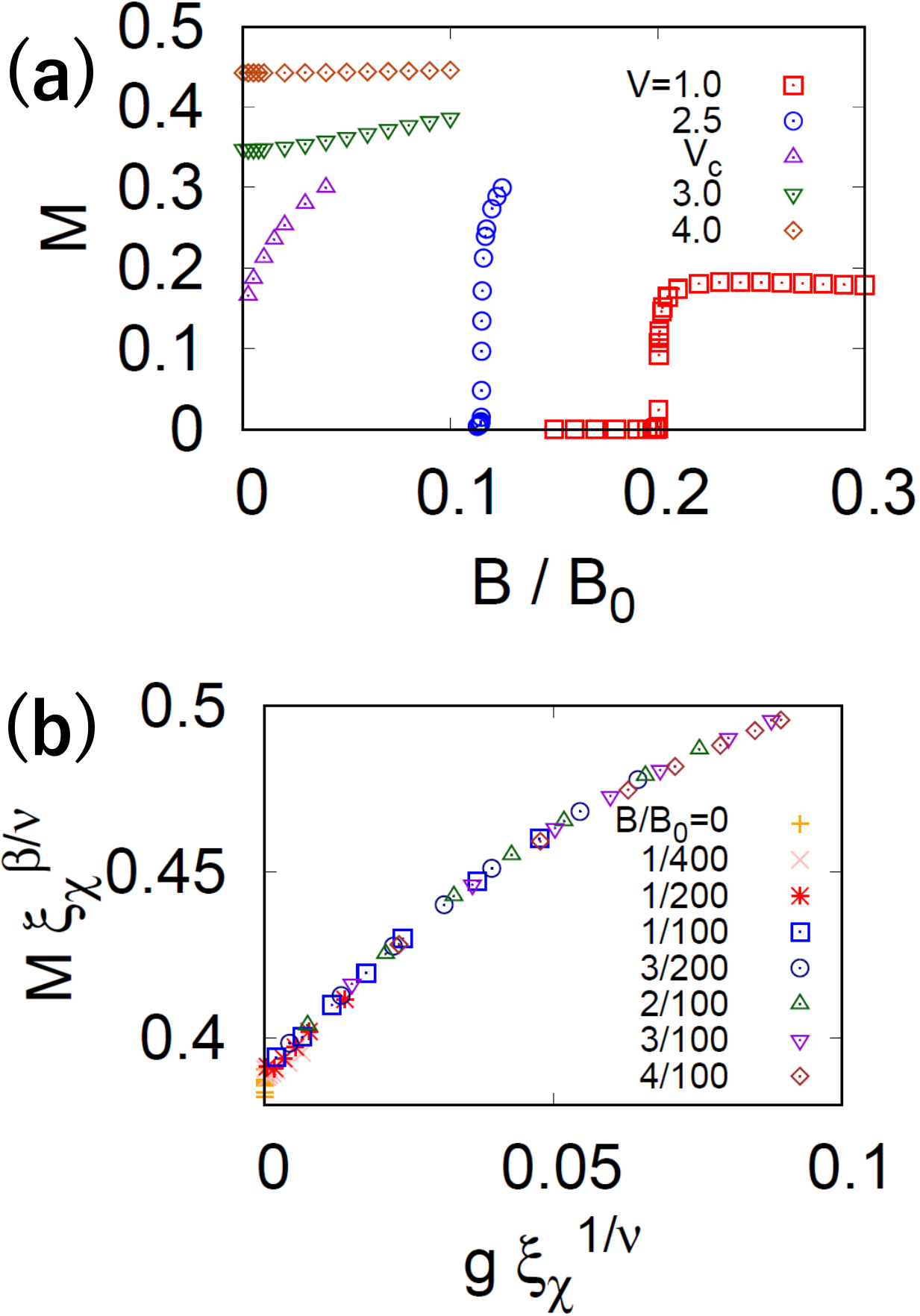}
\caption{
(a) The CDW order parameter $M$ for several values of the interaction $V$ calculated by iDMRG
with the fixed bond dimension $\chi=200$.
(b) The scaling plot of $M$ at the critical point $V=V_c=2.8678t$
in terms of the reduced coupling $g=(B/B_0)^2$ and the correlation length due to the finite bond dimension 
$\xi_{\chi}$.
The critical exponents are fixed as $\beta=0.125$ and $\nu=1$.
The bond dimensions used in the calculation are $\chi=20\sim 200$. 
}
\label{fig:L2}
\end{figure}

Figure~\ref{fig:L2}(a) shows the CDW order parameter $M$ as a function of the magnetic field $B$ for various 
interaction strengths.
Differently from the cylinder geometry discussed in the main text,  
$M$ remains zero for some range of  $B\neq0$ when the interaction $V$ is smaller than the critical value
$V_c(L_y=2)=2.8678t$.
The field induced phase transitions for $V<V_c(L_y=2)$ are so sharp that it is difficult to numerically 
identify the nature of these phase transitions.
On the other hand, $M(B)$ at the critical point exhibits the conventional $(1+1)$-dimensional Ising criticality
with the critical exponents $\beta=0.125,\nu=1$ as shown in Fig.~\ref{fig:L2}(b).
Here, the reduced coupling constant is chosen as $g=(B/B_0)^2$ and the cut-off length scale is
the correlation length due to the finite bond dimension $\xi_{\chi}$ computed from the second largest
eigenvalue of the transfer matrix.
The Ising criticality is consistent with the previous study for the two-leg ladder 
with no magnetic field whose criticality is described by 
free Marajona fermions~\cite{Tada2019}.

\begin{thebibliography}{70}
\bibitem [{MF()}]{MF}%
  \BibitemOpen
  \href@noop {} {}\bibinfo {note} {In the mean field calculation, the
  interaction term is approximated as $n_in_j\rightarrow n_i\langle
  n_j\rangle+\langle n_i\rangle n_j-\langle n_i\rangle\langle
  n_j\rangle$.}\BibitemShut {Stop}%
\bibitem [{hyp()}]{hyperscaling}%
  \BibitemOpen
  \href@noop {} {}\bibinfo {note} {The order parameter anomalous dimension is
  $\eta=1$ in the large $N_f$ limit, and therefore the hyperscaling relation
  holds within mean field approximations in the present $D=3$ system. This is
  consistent with the result that our scaling ansatz with respect to $l_B$
  holds at the mean field level.}\BibitemShut {Stop}%
\end{thebibliography}

%

\end{document}